\newcommand{\rem}[1]{}
\documentclass[prl,twocolumn,showpacs,preprintnumbers,amsmath,amssymb,floatfix]{revtex4}
\usepackage{hyperref}
\usepackage{epsfig}

\begin{document}

\title{Static friction on the fly: velocity depinning transitions of lubricants in motion}

\author{A. Vanossi$^1$, N. Manini$^{2,3}$, F. Caruso$^{2}$,
        G.E. Santoro$^{3,4}$, and E. Tosatti$^{3,4}$}
\affiliation{$^1$CNR-INFM National Research Center S3 and Department of Physics, \\
University of Modena and Reggio Emilia, Via Campi 213/A, 41100 Modena, Italy \\
$^2$Dipartimento di Fisica and CNR-INFM, Universit\`a di Milano, Via Celoria 16, 20133 Milano, Italy \\
$^3$International School for Advanced Studies (SISSA)
and CNR-INFM Democritos National Simulation Center, Via Beirut 2-4, I-34014 Trieste, Italy \\
$^4$International Centre for Theoretical Physics (ICTP), P.O.Box 586, I-34014 Trieste, Italy
}

\date{\today}

\begin{abstract}
The dragging velocity of a model solid lubricant
confined between sliding periodic substrates exhibits a
phase transition between two regimes, respectively with quantized
and with continuous lubricant center-of-mass velocity.
The transition, occurring for increasing external driving force
$F_{\rm ext}$ acting on the lubricant, displays a large
hysteresis, and has the features of depinning transitions in
static friction, only taking place on the fly.
Although different in nature, this phenomenon appears isomorphic
to a static Aubry depinning transition in a Frenkel-Kontorova model, the
role of particles now taken by the moving kinks of the
lubricant-substrate interface.
We suggest a possible realization in 2D optical lattice experiments.
\end{abstract}


\maketitle


One of the current areas of development in the modern understanding of
sliding friction is the depinning transition between substrate and slider
in static friction\cite{Rubinstein04}.
A slider initially pinned on a substrate requires a finite force in order
to move, whereas an unpinned one will move even under an infinitesimal
force.
It is generally known that ideal crystalline sliders commensurate
with periodic substrates are always pinned, whereas incommensurate
ones exhibit as a function of material parameters a sharp
transition between a rigid, pinned state for soft sliders on
strongly attractive substrates and a freely sliding state for hard
sliders on weakly attractive perfect substrates. The standard
paradigm for that transition \cite{Shinjo93} is the Aubry
transition in the time-honored one-dimensional (1D) Frenkel-Kontorova (FK) 
model \cite{braunbook,Aubry83}.
Once sliders are set into motion, however, one would generally
expect no further rigidity or pinning to persist or to appear. It
was thus rather surprising to find that simple systems like the
3-lengthscale driven 1D model of Ref.~\cite{Braun05}, inspired by
the tribological problem of two sliding crystal surfaces with a
thin solid lubricant layer in between, exhibits a novel kind of
rigidity, now around a state of dynamical motion, where the
lubricant center-of-mass (CM) velocity is robustly locked onto
quantized plateau values
\cite{Vanossi06,Santoro06,Manini07extended}.
The dynamical pinning of the lubricant's motion onto a rigidly quantized
relative velocity state has been understood in terms of kinks of the
lubricant being set into motion by the shear due to the moving
surfaces.
This kind of kink dragging can be argued to represent a rather general
mechanism, possibly at play also in more realistic two-dimensional
situations.

%
One such situation could be realized e.g., by attempting to slide
a rare gas physisorbed layer on a crystalline surface, by means of an
external dragging agent, such as an AFM tip, or an optical tweezer, or
a 2D optical lattice. In addition, the substrate could be
oscillated in a Quartz Crystal Microbalance (QCM), 
the inertial force acting as an (infinite wavelength) dragging agent.
In either case, the external potential would
in reality drag the {\em solitons} or discommensurations formed by
the adsorbate with the substrate -- it would drag the 2D Moir\'e patterns.
While such a class of phenomena has not yet been explored, it is
potentially quite interesting, and its extent and consequences deserve
a full theoretical understanding.
In this Letter we present a close analysis revealing that the
lubricant velocity-pinned state shares many more of the characters
typical of ordinary pinning in static friction, including hysteresis
against a depinning external force $F_{\rm ext}$, and 
a genuine depinning transition as a function of lubricant stiffness, that
appears a dynamical isomorph to the Aubry transition of static friction.

Starting with a chain of harmonically interacting particles \cite{harmonic:note}
(the lubricant layer), the Hamiltonian \cite{Vanossi06,Santoro06}
\begin{eqnarray} \label{hamiltonian:eq}
H &=&\sum_i
\Big[\frac{ p_i^2}{2m} +  \frac K2 (x_{i+1}-x_i-a_0)^2
\\ \nonumber
&+&\!\!\frac{U_+}{2} \cos{\frac{2\pi\,x_i}{a_+}
} + \frac{U_-}{2} \cos{\frac{2\pi\,(x_i-v_{\rm ext}t)}{a_-} }
-F_{\rm ext}\,x_i \Big]
\end{eqnarray}
contains the interaction with the two mutually sliding substrates of
sinusoidal amplitudes $U_+$ and $U_-$ and periods $a_+$ and $a_-$, 
different from the rest length $a_0$ of the springs, which sets a third length scale.
Introducing dimensionless length ratios $r_{\pm}=a_{\pm}/a_0$ we
assume $r_->\max(r_+,r_+^{-1})$, whereby the $(+)$ slider has the
closest periodicity to the lubricant, the $(-)$ slider the
furthest.
To study depinning, we here apply a uniform external force
$F_{\rm ext}$ to all chain particles.
The infinite chain size is managed -- in the general
incommensurate case -- by means of periodic boundary conditions
(PBC) and finite-size scaling \cite{Vanossi06,Manini07extended}.
Sliding the substrates at constant relative velocity $v_{\rm ext}$, 
the lubricant equations of motion are integrated by a
standard Runge-Kutta algorithm, including a phenomenological
viscous friction force
$
-2\gamma\,(\dot{x}_i - \frac 12 v_{\rm ext})=
-2\gamma\,(\dot{x}_i - v_{\rm w})
$
accounting for various sources of
dissipation. As shown earlier \cite{Vanossi06,Santoro06},
for $F_{\rm ext}=0$ and a wide
range of parameters, the sliding chain reaches a dynamical
stationary state characterized by a {\em quantized} plateau velocity
\begin{equation}\label{vplateau}
v_{\rm cm}/v_{\rm ext} =
v_{\rm plateau}/v_{\rm ext} \equiv 1 - r_+^{-1}\,,
\end{equation}
depending solely on the incommensurability ratio $r_+$, independent
of $r_-$, $K$, $\gamma$, $v_{\rm ext}$, and even of $U_-/U_+$.
In this quantized state of motion the array of
topological solitons (kinks for $r_+>1$ or antikinks for $r_+<1$) formed by
the chain with the least mismatched $(+)$ substrate is rigidly dragged at
velocity $v_{\rm ext}$ by the most mismatched $(-)$ substrate.
This dynamical pinning of the kink velocity to the moving $(-)$ slider
suggests a more complete analogy to the static pinned state of particles in
the standard FK model, where a finite force $F_{\rm ext}>F_c^{\rm FK}$ is
required to slide the chain over a periodic substrate.
Depending on the ratio $\theta^{\rm FK}$ between the substrate and chain
periodicities, static pinning in the FK model occurs for arbitrary
$K$ for rational $\theta^{\rm FK}$, whereas irrational $\theta^{\rm FK}$'s
show a depinning (Aubry) transition taking place at a critical
chain stiffness $K_{\rm Aubry}$, beyond which free sliding is induced by
arbitrarily small $F_{\rm ext}$ \cite{Aubry83,braunbook}.
In the present slider-lubricant-slider problem, the kinks formed
by the lubricant over the $(+)$ slider form a 1D lattice with an
average distance $d_{\rm k}=a_+/(r_+ -1)$, implying a kink
coverage $\theta=a_-/d_{\rm k}=r_-\,(1-r_+^{-1})$ of the $(-)$
substrate (for antikinks, $\theta<0$). If the kinks here play the
role of the particles in the FK model, we should be able,
depending on the relative commensurability $\theta$,
to observe either indefinite pinning (i.e., quantized lubricant velocity) in the
commensurate case, or else an Aubry-like depinning transition in
the incommensurate case. We find that this is what happens,
moreover with a clearly outlined isomorphism of the dynamical kink
pinning problem to the static particle pinning of the FK model.

\begin{figure}
\epsfig{file=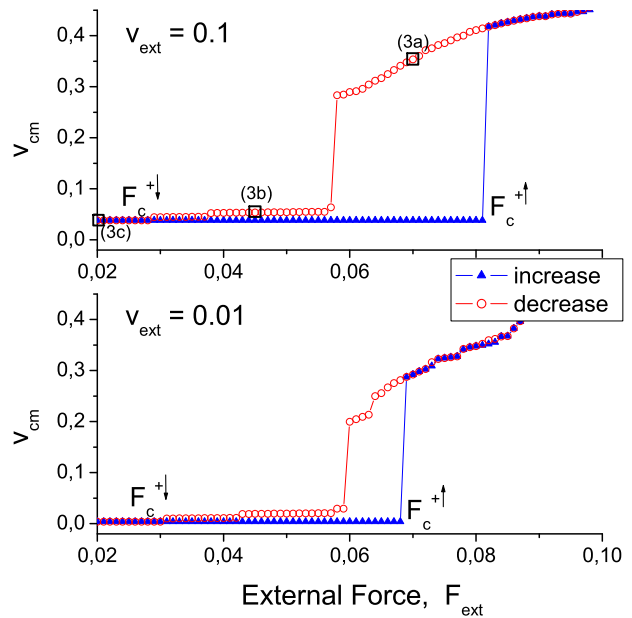,width=82mm,angle=0,clip=}
\caption{\label{hyst_loop}
(Color online) Hysteresis in the $v_{\rm cm}-F_{\rm ext}$ characteristics
for $(r_+,r_-)=(\phi,\phi^2)$.
Throughout, our units are $m=1$, $a_+=1$, $F_+=2\pi U_+/a_+=1$, 
$K$ is measured in $F_+/a_+$, $\gamma$ in $\sqrt{mF_+/a_+}$, velocities in $\sqrt{a_+F_+/m}$,
and we choose $F_-=2\pi U_-/a_-=F_+$. \cite{Manini07extended}
The behavior is shown for fast ($v_{\rm ext}=0.1$, upper panel) and slow
($v_{\rm ext}=0.01$, lower panel) drive.
Adiabatic increase and decrease of $F_{\rm ext}$ is denoted by triangles
and circles, respectively. A characteristic multi-step feature appears when
decreasing adiabatically $F_{\rm ext}$.
Here $\gamma=0.1$ and $K=4$.
}
\end{figure}

In order to highlight this deep similarity with the static case,
beginning from a quantized-velocity state, we start off by
studying the motion of the chain under the action of an
adiabatically cycled external force $F_{\rm ext}$.
We select two prototypical kink coverages,
one where the kink lattice is commensurate with the $(-)$ dragging
slider ($\theta=1$, realized by $r_+=\phi$, $r_-=\phi^2$, with
$\phi\equiv (1+\sqrt{5})/2$, the golden mean), and another
incommensurate ($\theta=\phi$, obtained with $r_+=\phi$,
$r_-=\phi^3$). The exactly quantized CM velocity plateau at
$F_{\rm ext}$ = 0 suggests null response to 
perturbations, so that any $0 < F_{\rm ext} < F_c^{+\uparrow}$
should have no effect whatsoever on $v_{\rm cm}$
(``incompressibility''). 

\begin{figure}
\epsfig{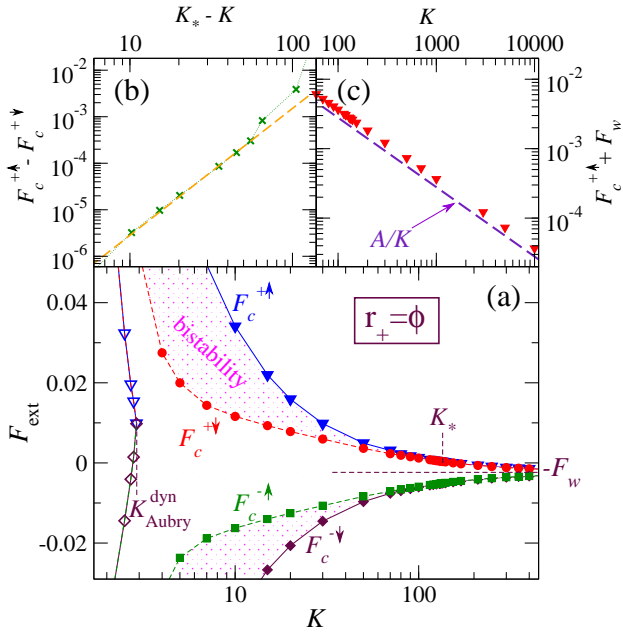}
\caption{\label{criticalF}
(Color online)
({\bf a}) The critical depinning ($F_c^{+\uparrow}$,
$F_c^{-\downarrow}$) and re-pinning ($F_c^{+\downarrow}$,
$F_c^{-\uparrow}$) forces, as functions of the chain stiffness
$K$, for $(r_+,r_-)=(\phi,\phi^2)$ (i.e.\ {\em commensurate}
$\theta=1$, solid symbols, pinning extending to infinitely large
$K$) and for $(r_+,r_-)=(\phi,\phi^3)$ (i.e.\ {\em incommensurate}
$\theta=\phi$, open symbols). 
In all calculations $v_{\rm ext}=0.1$, $\gamma=0.1$. 
For $\theta=1$:
({\bf b}) the width $F_c^{+\uparrow} - F_c^{+\downarrow}$ of the
bistability region as a function of $(K_*\!-\!K)$ (crosses), compared to a
fitted power-law
$F_c^{+\uparrow} - F_c^{+\downarrow} = B\,(K_*-K)^\alpha$,
with $\alpha\simeq 2.7$, $K_*\simeq 135$;
({\bf c}) the large-$K$ depinning force, compared to the expected power-law
$(F_c^{+\uparrow} + F_{\rm w}) = A\,K^{-1}$, Eq.~\eqref{largeK}.
}
\end{figure}

Figure~\ref{hyst_loop} displays the result of cycling the external
force $F_{\rm ext}$ up and down in small steps, for $\theta=1$,
$\gamma\,(K\,m)^{-1/2}\ll 1$ (underdamped regime), two different $v_{\rm ext}$,
and a chain stiffness $K$ within the velocity plateau. 
It displays a clear incompressibility and a hysteretic loop, analogous to the
depinning transition in static friction \cite{Rubinstein04}, and
in the FK model \cite{braunbook}.
For increasing $F_{\rm ext}$ the lubricant average velocity $v_{\rm cm}$ is
discontinuous at $F_c^{+\uparrow}$, with a {\em dynamical depinning}, a
finite jump $\Delta v$ and a bistable behavior. When $F_{\rm ext}$ is
decreased back, the depinned state survives down to $F_c^{+\downarrow} <
F_c^{+\uparrow}$, where quantized sliding is retrieved, as shown in
Fig.~\ref{hyst_loop}.
The hysteretic behavior observed under the application of a
negative $F_{\rm ext}$ is similar to that for $F_{\rm ext}>0$,
with corresponding $F_c^{-\downarrow} < F_c^{-\uparrow}<0$.
Viewing the kinks of our model as the particles of a standard FK
model, the $\theta=1$ case is isomorphic to the fully commensurate
$\theta^{\rm FK}=1$ FK static case, where pinning holds for
all $K$'s.
Confirming that, when we counterbalance the average frictional
force $F_{\rm w}=-2 \gamma\,(v_{\rm plateau}-v_{\rm w})$ with an
equal and opposite external force $F_{\rm ext}=-F_{\rm w}$ such
that no net force acts on the kink lattice, we
find that the plateau extends out to arbitrarily large $K$, see
Fig.~\ref{criticalF}(a).
%
The observed phenomenology is quite generic:
in suitable parameters ranges we find
similar hysteretic plateaus for all values of $r_+$ and $r_-$
investigated.
Moreover, different choices of $r_- = \theta/(1-r_+^{-1})$, with
rational $\theta\neq 1$ lead to plateaus which also extend to arbitrarily
large $K$.
A generic rational $\theta>1$ implies defects in the perfectly commensurate
($\theta=1$) lattice of kinks, in the form of kinks of the kink lattice
--- hierarchical excitations, {\em kinks of kinks} --- which, being
commensurate with $(-)$ substrate, resist depinning for arbitrarily large $K$.
%

On the contrary,
we expect --- and find, see Fig.~\ref{criticalF}(a), for $\theta=\phi$ --- 
that irrational
$\theta$ should be associated with a genuine Aubry transition {\em
on the fly} of the incommensurate lattice of kinks, with the
plateau disappearing at $K_{\rm Aubry}^{\rm dyn}$. 
The transition takes place at a finite $K=K_{\rm Aubry}^{\rm dyn}$,
where the width $F_c^{+\uparrow} - F_c^{-\downarrow}$ of the velocity
plateau as a function of $F_{\rm ext}$ shrinks to zero.
%
%
Of course, we expect $K_{\rm Aubry}^{\rm dyn}$ to be a
complicated function of the original model parameters, generally not
coincident with the value characteristic of the FK model for
$\theta^{\rm FK}=\theta$.

The pinning force $(F_c^{+\uparrow} + F_{\rm w})$ decreases like
$K^{-1}$ for large $K$, see Fig.~\ref{criticalF}(c).
In this limit, the kink lattice becomes increasingly faint, since
the corresponding particle-density modulation amplitude drops as
$K^{-1}$.
Quantitatively, the small displacements of the particles away from ideal
lattice positions $x_i=i a_0$ due to interaction with the $(+)$ potential
are described in terms of a hull function, whose explicit expression is
known \cite{vanErp99} for large $K\gg F_+/a_+$ (with $F_+=2\pi\, U_+/a_+$).
These displaced positions, once substituted into the $U_-$ term of
Eq.~\eqref{hamiltonian:eq}, yield a sinusoidal oscillation as a function of
the translation $v_{\rm ext}t$ of the $(-)$ substrate relative to the kink
lattice.
This energy oscillation corresponds to a force whose maximum amplitude
equals the minimum force that must be applied to the chain to have the
kinks overcome the barrier and initiate sliding.
This critical force per particle amounts to
\begin{equation} \label{largeK}
F_c^+ + F_w = \frac AK + O(K^{-2}), \quad
A = \frac{\pi}8\, \frac{r_+ -1}{\sin^2\!\frac \pi {r_+}}\, \frac{F_+F_-}{a_+}
\,,
\end{equation}
and is drawn in Fig.~\ref{criticalF}(c) for comparison.
The $\simeq 20\%$ discrepancy with the observed $F_c^+$ is due to the
neglect of the displacements induced by the $(-)$ potential, which
enhance the kinks amplitude in this $\theta=1$ geometry. 
%

\begin{figure}
\epsfig{file=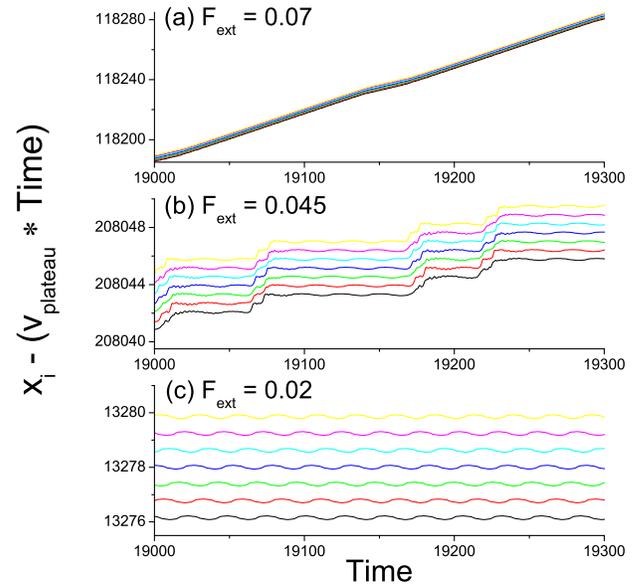,width=82mm,angle=0,clip=}
\caption{\label{trajectories:fig}
(Color online) Time evolution of the positions $x_i - v_{\rm plateau}\cdot t$ 
of 7 contiguous chain particles. 
$t$ is measured in units of $\sqrt{ma_+/F_+}$.
The plots refer to three dynamical regimes observed along the
adiabatic decrease of $F_{\rm ext}$, at the points indicated by (3a)-(3c)
in Fig.~\ref{hyst_loop}:
(a) free sliding at $F_{\rm ext}=0.07$,
(b) dynamic stick-slip at $F_{\rm ext}=0.045$, and
(c) quantized sliding state at $F_{\rm ext}=0.02$.
Same parameters as in Fig.~\ref{hyst_loop}, with $v_{\rm ext}=0.1$.
}
\end{figure}

The self-pinning at the $(-)$ minima is also at the root of the bistability
found at moderate $K$: self-enhanced trapped kinks resist to a large
$F_{\rm ext}$ up to $F_c^{+\uparrow}$, but as soon as kinks wash away in
the large-$F_{\rm ext}$ unpinned sliding state, $F_{\rm ext}$ must be
reduced to smaller $F_c^{+\downarrow}$ before kinks reconstruct and then
re-pin.
The transition boundaries $F_c^{+\uparrow}$, $F_c^{+\downarrow}$, and
$\Delta v$ are nontrivial functions of the parameters.
In particular, as shown in Fig.~\ref{criticalF}(a), for large $K$ the
hysteretic behavior extends up to $K=K_*\simeq 135$, see
power-law fit in Fig.~\ref{criticalF}(b).
For $K\geq K_*$, $F_c^{+\uparrow}\equiv F_c^{+\downarrow}=F_c^+$,
and the depinning transition is continuous, characterized by a
critical behavior $(v_{\rm cm}-v_{\rm plateau})\propto(F_{\rm ext}-F_c^{+})^{1/2}$. 
On approaching the critical point $K_*$, the
jump $\Delta v$ characterizing the first-order depinning
transition vanishes, and a continuous transition line originates
there.
In the moderate-$K$ hysteretic regime we find that, in a wide
range of parameters, depinning occurs through a
discontinuous jump to a quasi-free sliding regime $F_{\rm ext} > F_c^{+\uparrow}$, 
see Fig.~\ref{trajectories:fig}(a),
characterized by aperiodic single-particle motion, superposed to
a drift approaching the
free-particle limit velocity $F_{\rm ext}/(2\gamma)$. 
The reverse route, 
upon decreasing $F_{\rm ext}$ to $F_c^{+\downarrow}$, yields some
insight in the re-pinning mechanism.
Depending on the values of $K$ and $v_{\rm ext}$, dynamical
re-pinning to the plateau-state takes place through a
variety of mechanisms, from intermittencies with a well defined
periodicity, to more chaotic and irregular jumps.
Despite that variety, for intermediate $K$ we
consistently observe one or more multi-step sliding regimes as
$F_{\rm ext}$ decreases within the $[F_c^{+\downarrow},F_c^{+\uparrow}]$ hysteresis window,
consistently with the results known for the sine-Gordon \cite{Bishop87}, and FK \cite{Braun97}
models.
%
Our intermediate-$F_{\rm ext}$ regime, illustrated in
Fig.~\ref{trajectories:fig}(b), is reminiscent of intermittencies
representing a {\em dynamic stick-slip} occurring on a local scale, clearly seen
by plotting the particle trajectories in the
reference frame which slides at $v_{\rm plateau}$.
In fact, detailed global analysis of this intermediate-$F_{\rm ext}$ 
dynamics, based on simulations with
$r_+=1+\epsilon$ (with $\epsilon\ll 1$) where kinks are
well separated, shows that the kinks pin to the $(-)$ lattice
along most of the chain, with a few kink-antikink defects
travelling along the chain.
These defective sections move along the chain at 
a characteristic speed depending on the model parameters.
%
The nature of these de-pinning mobile
defects is reminiscent of those found in the ordinary FK model
\cite{braunbook}. Depending on $K$ and $F_{\rm ext}$, single- or
multiple-kink defects arise: transitions between intermediate
phases characterized by different numbers of travelling-kink
defects are visible in Fig.~\ref{hyst_loop}.

\begin{figure}
\epsfig{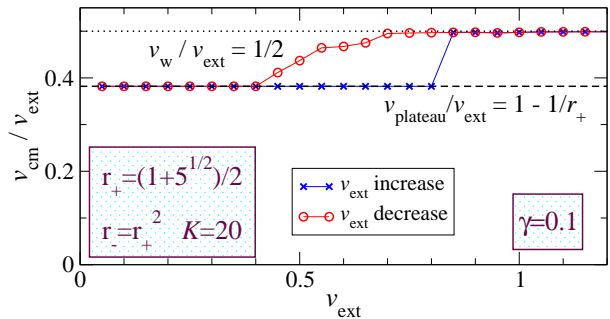}
\caption{\label{Vloop_hyst:fig}
(Color online)
Hysteresis, as $v_{\rm ext}$ is cycled, in the underdamped regime.
}
\end{figure}

In practice, the experimental realization of an equal force
acting on each particle could be provided, for instance, by the inertial forces
in a QCM experiment.
%
%
In addition, the hysteretic depinning and re-pinning occurs even when different tunable 
parameters are varied adiabatically, for instance the driving velocity $v_{\rm ext}$,
see Fig.~\ref{Vloop_hyst:fig}, as long as we are in the underdamped limit \cite{overdamped:note}.
%
%
Thus, in a concrete lab experiment, cycling quantities such as $v_{\rm ext}$
should lead to leaving/recovering the plateau state, similar to cycling $F_{\rm ext}$.
As for the dragging agent, the typical force that a laser provides in an optical lattice
of alkali atoms is $F\approx 10^{-9}$pN \cite{Bloch}. 
This is very small, but we checked that our phenomenology still survives even when $F_-$ is orders of
magnitude smaller than $F_+$.
Larger (inertial) forces, $m\omega^2 a \sim 10^{-5}$pN per adsorbate atom, are accessible in 
a standard QCM oscillating at $\omega/2\pi \sim 15$ MHz with amplitudes $a\sim 100$ \AA. 
The sinusoidal variation of the dragging velocity occurs here between $0$ and $\pm 1$m/s on a much longer time-scale
than the typical adsorbate vibrational frequencies ($\sim 1$ THz).
The inertial dragging force couples to the local density fluctuations, 
and should drag the solitons exactly like a commensurate $(-)$ slider would do. 

%
As for thermal effects, they favor unpinning, but as long as $k_BT$
is much smaller than the cost $\Delta$ for the creation of the relevant
depinning defects, the physics does not change much, as shown by numerical
simulations of similar models.\cite{Braun97PRE} 
%
%

Our overall picture confirms a striking similarity of the
dynamical kink-depinning transition of a lubricant under shear to
the usual depinning in static friction, and in the static FK
model, to which the large-$K$ theory of Eq.~(\ref{largeK})
establishes a quantitative correspondence. 
These findings should also be amenable to experimental verification.
%
%

\begin{acknowledgments}
This research was partially supported by PRRIITT (Regione Emilia
Romagna), Net-Lab ``Surfaces \& Coatings for Advanced Mechanics
and Nanomechanics'' (SUP\&RMAN) and by MIUR PRIN 2006022847.
\end{acknowledgments}


\begin{thebibliography}{10}
\bibitem{harmonic:note}
{The choice of a harmonic interaction is not crucial: simulations with
  Morse-type and Lennard-Jones potentials also reveal the ubiquity of the
  observed phenomenology.}
\bibitem{overdamped:note}
{ In the overdamped limit, when the viscous rate $\gamma/m$ is larger
  than the characteristic frequencies of the kink-lattice, the
  dynamical depinning turns second order. }
\end{thebibliography}

\end{document}